\documentstyle[12pt,epsf]{article}
\large

\def\la{\mathrel{\mathpalette\fun <}}
\def\ga{\mathrel{\mathpalette\fun >}}
\def\fun#1#2{\lower3.6pt\vbox{\baselineskip0pt\lineskip.9pt
\ialign{$\mathsurround=0pt#1\hfil##\hfil$\crcr#2\crcr\sim\crcr}}}

\newcommand{\bc}{\begin{center}}
\newcommand{\ec}{\end{center}}
\newcommand{\bd}{\begin{displaymath}}
\newcommand{\ed}{\end{displaymath}}
\newcommand{\be}{\begin{equation}}
\newcommand{\ee}{\end{equation}}
\newcommand{\ba}{\begin{array}}
\newcommand{\ea}{\end{array}}
\newcommand{\bt}{\begin{tabular}}
\newcommand{\et}{\end{tabular}}

\begin{document}

~\hfill ITEP-96-8

\begin{center}
\large
THE STRUCTURE FUNCTIONS OF LONGITUDINAL VIRTUAL \\
PHOTON AT LOW VIRTUALITIES 
\end{center}

\begin{center} {\bf
B.L. Ioffe} and {\bf I.A. Shushpanov}
\end{center}
\begin{center}
\it{Institute for Theoretical and Experimental Physics, 
B. Cheremushkinskaya 25, Moscow 117259, Russia}
\end{center}

\vspace{1.cm}
\centerline{December 1995}
\vspace{1.5cm}
\begin{abstract}

The structure functions $F^L_1$ and $F^L_2$ of longitudinal virtual photon 
at low virtualities are calculated in the framework of chiral pertubation
theory(ChPT) in the zero and first order of ChPT. It is assumed that
the virtuality of target longitudinal photon $p^2$ is much less 
than the virtuality of the hard projectile photon $Q^2$ and both 
are less than the characteristic ChPT scale. In this 
approximation the structure functions are determined by the production
of two pions in $\gamma \gamma$ collisions. The numerical results for
$F^L_2$ and $F^L_1$ are presented (the upper index refers to longitudinal
polarization of the virtual target photon). The possibilities of
measurements of these structure functions are briefly discussed. 

\end{abstract}

\section{ Introduction}

$\quad$ As was mentioned in Ref's[1,2] the structure function of longitudinal
virtual photon $F_2^L(x,p^2)$ in massless QCD is nonvanishing, when
photon virtuality $p^2$ is going to zero. This unusual circumstance
is caused by the fact that the imaginary part of the forward 
$\gamma \gamma$ scattering amplitude (box diagrams with quarks in 
the loops), determining the photon structure function , has a pole
 $1/p^2$ in the virtuality of the target photon. After multiplication
by the product of longitudinal photon polarizations in the initial
 and final states, proportional to $p^2$ $(e^L_i e^L_f \propto p^2)$,
this results in nonvanishing $F_2^L(x,p^2)$ at $p^2 \rightarrow 0$.

 The seemingly nonvanishing interaction of longitudinal
photons at $p^2 \rightarrow 0$ in massless theories does not result[3,4] in
any inconsistency of the theory: the arising problems in the gedanken
experiments with longitudinal photons are solved by exploiting the concept
of the formation zone[3]. There are nonpertubative effects which
are important for determination of the longitudinal photon structure
function $F_2^L(x,p^2)$ at low $p^2$. For this reason it is impossible
to find it basing on QCD now. Even the approach, using the operator
product expansion in $1/p^2$ and the analytical continuation in the 
domain of low $p^2$, which was succeful in the determination of structure
function of real and virtual transverse photon[2,5] is not working here:
the extrapolation to low $p^2$ appears to be impossible.

 At low virtualities and energies of all initial and final particles
in the given process QCD is equivalent to the ChPT: the theory 
of almost massless particles - pions (for reviwes see[6-8]).
 Therefore we may expect
that the peculiar properties of longitudinal photon in massless QCD will
manifest themselves in ChPT with massless pions. This idea was proposed
in Ref.[1].
 
 The goal of this paper is to perform the quantitave calculations of
the longitudinal photon structure functions at low $p^2$ basing on ChPT.
In order to be in the framework of ChPT it is necessary to assume 
that not only virtuality of target (soft) photon $p^2$ is less than the
applicability limit of ChPT $M^2 \approx m_{\rho}^2 = 0.6$GeV$^2$, 
but also that the virtuality of the probe (hard) photon $Q^2=-q^2$ in
$\gamma \gamma$ scattering is below this limit. It means that the 
photon structure functions are not in the scaling region
of $Q^2$. 

 There is the essential difference in the photon structure functions
at high and low virtualities $Q^2$ of probe (hard) photon. At high $Q^2$
the structure functions which are transverse in the polarization of
the hard photon are dominating for the case of 
 longitudinal, as well as for 
transverse polarized soft photon. This is a direct consequence of 
the fact that at high $Q^2$ the $\gamma \gamma$ scattering proceeds 
through the $\gamma$-scattering on quarks. At low $Q^2$ and $p^2$
the $\gamma \gamma$ scattering cross section is determined by the production 
of pions. It is known [9], that in the virtual photon scattering on free
spinless boson the cross section of the longitudinally polarized
photon is dominating. This means, that at low $Q^2$ and $p^2$ the 
longitudinal structure function will exceed the transverse one, 
$F_L\approx F_2/2x \gg F_1$.

In this paper the longitudinal virtual photon structure functions are
calculated in the zero and first order in ChPT. The results are presented
for structure functions $F^L_L(x,p^2,Q^2)$, $F^L_2(x,p^2,Q^2)$
 and $F^L_1(x,p^2,Q^2)$,
where the upper index refers to the polarization of the soft (target)
photon with momentum $p$ and the lower index refers to the polarization
of the hard (projectile) photon with momentum $q$, $-q^2=Q^2$.

 The  expansion parameter in ChPT is $M^2=C\pi^2 F_{\pi}^2$, where 
$F_{\pi}=93$ Mev is the pion decay constant and $C$ is the constant 
depending on process and in this case it is about~6. Let us assume that
$Q^2/M^2$, $p^2/M^2$ and $p^2/Q^2$ are the parameters of expansion
 and restrict ourselves by the first
order terms in these ratios. For the value of $s=(p+q)^2$ the more weak
limitation is imposed: $s \la M^2$ because $s$ explicitly enters only in 
the small corrections to the results.
The pion mass $\mu$ will be partly accounted:
no assumption $|p^2| \gg \mu^2$ is used but corrections proportional to
$\mu^2/Q^2$ and $\mu^2/s$ are neglected, besides the overall phase 
space factor.

 In the section 2 the results of calculations of $F^L_2$, $F^L_1$
 and $F^L_L$ in 
the zero order of ChPT are presented. In section 3 the first order terms
in ChPT are calculated. In the section 4 the numerical results for the 
structure functions are given and the possibilities
of the measurements of these structure functions are briefly discussed.  
 
\section{Calculation of the Structure Functions in the zero order
of ChPT}

$\quad$ In order to get the structure function $F^L_2$ of the longitudinal
target
photon the imaginary part of  $\gamma \gamma$ forward scattering amplitude  
$ImT_{\mu\nu\lambda\sigma}(p,q)$ must be multiplied by the product of
longitudinal photon polarization in the initial and final states

\be
\label{polar}
e^L_{\lambda}(p)e^L_{\sigma}(p)=-\frac{p^2}{\nu^2-p^2 q^2}
(q_\lambda-\frac{\nu p_\lambda}{p^2})(q_\sigma-\frac{\nu p_\sigma}{p^2}),
\ee
where $\nu=pq$ and indeces $\mu,\nu$ and $\lambda,\sigma$ refer to
hard and soft photon correspondingly. In the final answer the terms
proportional to $p_\mu p_\nu$ must be separated

\be
\label{F2amp}
F^L_2(p^2,q^2,\nu)=2\alpha\nu\left\{ ImT_{\mu\nu\lambda\sigma}(p,q)
e^L_{\lambda}(p)e^L_{\sigma}(p)\right\}_{term\propto p_{\mu}p_{\nu}}
\ee
    
 In the zero order of ChPT the proportional to $p_{\mu}p_{\nu}$ term of 
imaginary part of  $\gamma \gamma$ forward
scattering amplitude $ImT_{\mu\nu\lambda\sigma}(p,q)$
 is determined by the two box diagrams Figs.~1a,b,
corresponding to standard scalar electrodynamics. It is easy to see
that the diagrams of Fig. 2, where two photon interact with two pions
at one point - the diagrams arising from $A_\mu^2\phi^+\phi$ term in 
Lagrangian of scalar electrodynamics do not contribute to ($\ref{F2amp}$)
because their contributions into imaginary part of amplitude have no terms
proportional to $p_{\mu}p_{\nu}$. (The contributions of the second terms 
in the brackets in ($\ref{polar}$) vanish due to current conservation
at the soft photon vertex.)

 The imaginary part of the amplitude is connected with discontinuity of
the amplitude in $s$ by:

\be
 ImT_{\mu\nu\lambda\sigma}(p^2,q^2,s)=\frac{1}{2i}
[T_{\mu\nu\lambda\sigma}(p^2,q^2,s+i0)
-T_{\mu\nu\lambda\sigma}(p^2,q^2,s-i0)].
\ee

 The calculation
of the diagrams of Fig.1 gives: for the direct diagram:

\be
\ba{ll}
\label{F21a}
F^{L(fig.1a)}_2=\frac{\alpha \chi}{2\pi}(1-4x^2\frac{p^2}{Q^2}) 
\left\{
x(1-2x)^2\left[(1-\frac{\mu^2}{p^2x(1-x+xp^2/Q^2)})^{-1}
(1+4\frac{p^2}{Q^2}x(1-2x))+\right.\right. \\[5mm]
\left.\left.+\frac{p^2}{Q^2}(-1+8x(1-x)+2x(2-3x)L)\right]
+2\frac{p^2}{Q^2}x(1-2x)[1+2x-6x^2-2x^2L]-
\frac{4}{3}x\frac{p^2}{Q^2}\right\}
\ea
\ee

and for the crossing diagram:

\be
\label{F21b}
F^{L(fig.1b)}_2=-\frac{\alpha \chi}{2\pi}\frac{p^2}{Q^2}x
\left\{(1-4x^2)[-1+x(1-x)(6+2L)]+\frac{2}{3}\right\}.
\ee

Here $x=Q^2/2\nu$ is the Bjorken variable, $\chi$ accounts two pion
phase space,
\be
\chi=\left(1-\frac{4\mu^2x}{Q^2(1-x)+p^2x}\right)^{1/2},
\ee
\be 
  L=ln\left[\frac{x^2}{Q^2}(-p^2+\frac{\mu^2}{x(1-x)})\right],  
\ee
and only the first order terms in the ratio $p^2/Q^2$ are retained.
As follows from ($\ref{F21a}$) in the theory with massless pions 
($\mu^2 \rightarrow 0$) the direct diagram results in nonvanishing
longitudinal photon structure function at $p^2 \rightarrow 0$, i.e.
for quasireal photon. In this aspect the situation in the ChPT resembles
the situation in the massless QCD[1,2]. In the real world  at
$p^2 \rightarrow 0$ $F^L_2$ is going to zero proportionally to $p^2$
starting from the values of $p^2$ of order $\mu^2/x(1-x)$. If $Q^2$ is
less than $M^2$, as we suppose, in this region of $p^2$ the terms,
proportional to $p^2/Q^2$ become of importance, as well as the terms
of the first order in ChPT.

 As is well known (see e.g. [9]), for any target the structure 
function $F_2$ is proportional to the sum of scattering cross sections
 $\sigma_{T}+\sigma_{L}$ of virtual transverse and longitudinal photons.
If the scattering proceeds on free massless fermion only $\sigma_T$
survives, $\sigma_L$=0. For the scattering on free massless boson the
situation is opposite: $\sigma_T=0$, $\sigma_L \neq 0$. Therefore, in the zero
order in ChPT the nonvanising contribution to $F^L_2$ comes from the
scattering of longitudinally polarized hard probe photon and it is reasonable
to calculate also $F^L_L(p^2,Q^2,x)$ - the structure function, corresponding
the case, where both photons are longitudinally polarized. The structure
function $F^L_L$ is related to $F^L_1$ and $F^L_2$ by:
\be
\label{FL}
F^L_L=-F^L_1+(1+4p^2x^2/Q^2)F^L_2/2x.
\ee

The structure function $F^L_L(p^2,Q^2,\nu)$ is expressed through the imaginary
part of the forward $\gamma\gamma$-scattering amplitude by:

\be
F^L_L(p^2,q^2,\nu)=2\alpha\left\{ ImT_{\mu\nu\lambda\sigma}(p,q)
e^L_{\lambda}(p)e^L_{\sigma}(p)e^L_\mu(q)e^L_\nu(q)\right\},
\ee
where $e^L_\mu(q)e^L_\nu(q)$ is given by ($\ref{polar}$) with substituion 
$\lambda\rightarrow\mu$, $\sigma\rightarrow\nu$ and $p\leftrightarrow q$.
 In the case 
of $F^L_L,$ unlike $F^L_2$, the diagrams of fig.2 are also contributing.

 Due to simple relation ($\ref{FL}$) between $F^L_2$, $F^L_1$ and  $F^L_L$
only results for $F^L_2$ and  $F^L_1$ will be represented.
The results of the calculation for $F^L_1$ in the scalar 
electrodynamics are the following:

\be
\label{FL1}
F^{L(scal.el.)}_1=\frac{\alpha \chi}{2\pi}\frac{p^2}{Q^2}x(1-x)
(1-2x)^2(L+3)
\ee 

  As is seen from ($\ref{FL1}$) in the $F^L_1$ there is not zero 
order term in the $p^2/Q^2$ expansion.

\section{The First Order ChPT Correction to Longitudinal Photon
Structure Functions}

 The general method of calculation of ChPT corrections is exposed in 
Ref's[6-8]. We will mainly use the results of Gasser and Leutwyler[6]
which are presented in the form most convenient for the present
calculations.

 From the general form of ChPT effective Lagrangian it can be easily 
shown that in the first order in ChPT only the two pions intermediate
states can contribute to the imaginary part of the forward $\gamma\gamma$-
scattering amplitude.

 To calculate corrections to the structure functions $F^L_L$ and $F^L_2$
in the first order of ChPT one should consider the general expression for
effective Lagrangian which contains all terms permited by chiral invariance
 in the order (momentum)$^4$.

 In the leading order of ChPT the effective Lagrangian has the form
\be
\label{ZO}
L^{(2)}=\frac{1}{2}F^2(\nabla_\mu U^T\nabla_\mu U)
\ee
where $U$ is a four-component ($U=\{U^0,U^i\}$) real $O(4)$ vector
of unit length $U^TU=1$. The space components of U are expressed 
through the pion fiels, $U^i=\phi^i/F$. The ChPT expansion corresponds
to the expansion in inverse powers of $F^2$. So, in the first
order of ChPT 
\be
U^0=1-\frac{1}{2}U^i U^i
\ee
The covariant derivative $\nabla_{\mu}$ is defined by
\be
\nabla_\mu U^0=\partial_\mu U^0
\ee
\be
\nabla_\mu U^i=\partial_\mu U^i+A_\mu\epsilon^{i3l} U^l,
\ee
where $A_{\mu}$ is the electromagnetic field. Expressing the real
pionic fields $\phi^1$ and $\phi^2$ through the fields of charged pions
$\phi^+$ and $\phi$
\be 
\phi^1=\frac{1}{\sqrt{2}}(\phi+\phi^+)\quad  
\phi^2=\frac{1}{\sqrt{2}i}(\phi-\phi^+)
\ee
and expanding ($\ref{ZO}$) up to terms $\sim1/F^2$, we have (only charged
pion fields are retained) 
\be
L^{(2)}=L^{scal.elec.}+\frac{1}{2F^2}\{\partial_\mu(\phi^+\phi)\}^2,
\ee
where
$L^{scal.elec.}=(\partial_\mu \phi+iA_\mu \phi)^+(\partial_\mu 
\phi+iA_\mu \phi)-
\mu^2 \phi^+ \phi$. Second term in $L^{(2)}$ corresponds to the four pion
interaction and leads to appearance of the loop corrections to the structure
functions which are proportional to $1/F^2$.

 In the order (momentum)$^4$, i.e. $\sim1/F^2$, there are two terms
in the general effective Lagrangian of ChPT[6], which are essential
for us
\be
L_5=l_5(U^T {\cal F_{\mu\nu} F_{\mu\nu}} U) 
\ee
\be  
L_6=l_6(\nabla_\mu U^T {\cal F_{\mu\nu}} \nabla_\nu U),
\ee
where 
\be
({\cal F_{\mu\nu}}U)^{i}=[(\nabla_\mu \nabla_\nu -\nabla_\nu\nabla_\mu)U]^{i}.
\ee
 In terms of
charged pionic fields 
\be
L_5=\frac{-2l_5}{F^2} F_{\mu\nu}^2 \phi^+ \phi
\ee
\be
L_6=\frac{-2il_6}{F^2}  F_{\mu\nu} \{ \partial_\mu \phi \partial_\nu 
\phi^+ +iA_\mu \partial_\nu(\phi^+ \phi) \},
\ee
where $F_{\mu\nu}$ is the electromagnetic field strength.
Therefore in this order the effective chiral Lagrangian is given by
\be
L=L^{scal.elec.} + \frac{1}{2F^2}\{\partial_\mu(\phi^+\phi)\}^2 +L_5+L_6,
\ee
where $L_5$ and $L_6$ are expressed in the terms of $\phi$,  $\phi^+$ 
and $A_\mu.$ $L_5$ and $L_6$ serve as the counter terms for the 
renormalization of loops: the infinities arising in loop calculations
are absorbed in $l_5$ and $l_6$, and as a result the finite values
$\bar{l_5}$ and $\bar{l_6}$ arise.

For the calculation of the first order
ChPT corrections to the structure functions we should calculate 
effective $\gamma\pi\pi$ and $\gamma\gamma\pi\pi$ vertices, 
substitute ones in all zero order diagrams of ChPT and collect terms
proportional to $1/F^2$. 

 At first let us consider effective $\gamma\pi\pi$ vertex. In the chosen
parameterization of U there are three diagrams which contribute to this
vertex in the first order of ChPT (Fig.3). First diagram corresponds
to vertex of $\gamma\pi\pi$ interaction in the scalar electrodynamics,
the second comes from $L_6$ and the third is the loop (unitary) diagram. 
 To renormalize the loop 
diagram it is convenient to use the dimensional regularization.
 The contribution
of this diagram to the $\gamma\gamma\pi$ vertex can be divided into
finite part and divergent part which contains the infinite factor 
$\lambda$, where $\lambda$ is :
$$\lambda=\frac{2}{4-d} + ln4\pi+1-\gamma_E - ln \frac{\mu^2}{\rho^2},$$
and $\rho$ is the scale of mass introduced by dimensional regularization.

 In the sum of the diagrams the divergent part will be absorbed into
$l_6$ and the following result for effective $\gamma\pi\pi$
vertex in the first order of ChPT was obtained[6]:
\be
-\frac{1}{i}\Gamma_\mu(k,k';q)=(k'+k)_\mu-\frac{\left(\bar{l}_6-1/3+
\sigma^2\{\sigma ln(\frac{\sigma-1}{\sigma+1})+2\}\right)}{48\pi^2F^2}
(q_\mu kq-k_\mu q^2),
\ee
where 
\be
\sigma=(1-4\mu^2/q^2)^{1/2},
\ee
$k$ and $k'$ are the pion initial and final momenta, q is the photon 
momentum, $k=k'+q.$ The numerical value of $\bar{l_6}$ was found in Ref. [6]
from the data on electromagnetic charge radius of the pion:
\be
\bar{l_6}=16.5\pm 1.1 
\ee
  After
substuting this effective vertex in zero order diagrams in all cases 
the following combination appears 

\be
\label{R6}
R_6(r^2)=\bar{l}_6-1/3+\sigma^2(r^2)\left\{\sigma(r^2)
ln\left[\frac{\sigma(r^2)-1}{\sigma(r^2)+1}\right]+2\right\}
\ee
where $r^2=q^2$ or $p^2$. The term proportional to $\sigma^2$ in ($\ref{R6}$) 
arises from the loop correction - the diagram Fig. 3c. It is much 
smaller numerically (about 10 times) than $\bar{l_6}$ and in what follows 
the difference between the values of $R_6(q^2)$ and $R_6(p^2)$
will be neglected. 
 
 In the case of $\gamma\gamma\pi\pi$ effective vertex there are six 
diagrams (fig. 4): first diagram comes from scalar electrodynamics,
second and third ones correspond to $L_6$ and $L_5$, respectively, and
the other diagrams are loop (unitary) corrections.

 In principle we can write out the tensor structure of this vertex
at once:
\be
\frac{1}{i}\Gamma_{\mu\lambda}(p,q)=2\delta_{\mu\lambda}-
\frac{R_6}{48\pi^2 F^2}[q_\lambda(p+q)_\mu-\delta_{\mu\lambda}(p+q,q)+
p_\mu(p+q)_\lambda-\delta_{\mu\lambda}(p+q,p)]+
B(p_\mu q_\lambda-pq \delta_{\mu\lambda} ).
\ee
 
 This equation follows from the fact that total amplitude for 
$\gamma\gamma\pi\pi$ scattering must contain the factor $R_6$ multiplyed
by tensor structure which is transverse in $q_\mu$ and $p_\lambda$
simultaneously. This conclusion can be checked by direct calculation.

 To simplify the calculation of this vertex we multiply 
$\Gamma_{\mu\lambda}$ by $ (p+q)_\mu(p+q)_\lambda$. After this 
procedure the term proportional to $R_6$ comes out. 
Taking into account that cutting pion propagators implies that
these pions are on mass-shell and substituting vertex for four-pion
interaction into diagrams 4(d-f), one can get the following
 expression for $B$ 

$$(B+\frac{8l_5}{F_{\pi}^2})(p^2q^2-\nu^2)=
\frac{(p+q)^2}{F^2}\frac{1}{i}
\int\frac{d^dk}{(2\pi)^d} \frac{ (p+q)^2+2k^2}
{(k^2-\mu^2)((k-p-q)^2-\mu^2)}-$$

\be
-\frac{1}{iF^2}\int\frac{d^dk}{(2\pi)^d}
\frac{\left\{2(k,p+q)+q^2+\nu\right\}
\left\{2(k,p+q)-p^2-\nu\right\}
\left\{(p+q)^2+(k+q)^2+(k-p)^2\right\}}
{\left(k^2-\mu^2\right)\left((k-p)^2-\mu^2\right)
\left((k+q)^2-\mu^2\right)}
\ee

As usual,
using dimensional regularization and absorbing the pole contributions,
 arising
from the loop diagrams,by infinite constant $l_5$ we obtain the
finite answer for B:
\be
B=\frac{\bar{l_5}-1/3+\sigma^2(q^2)\{\sigma(q^2)
 ln(\frac{\sigma(q^2)-1}{\sigma(q^2)+1})
+2\}+\delta }{24\pi^2F^2},
\ee
 where

$$\delta=Re\left[\frac{3}{2}(1-2x)\int^1_0dz\int^z_0dy \frac{s}
{\mu^2+y(1-y)Q^2-z(1-z)p^2-y(1-z)Q^2/x} -\right.$$

\be
\left.- 6(1-x)^2\left\{ \sigma(s) ln(\frac{\sigma(s)-1}{\sigma(s)+1})-
\sigma(q^2) ln(\frac{\sigma(q^2)-1}{\sigma(q^2)+1})\right\}\right].
\ee 
In the following the combination
$\bar{l_5}-1/3+
\sigma^2(q^2)\left\{\sigma(q^2)
 ln(\frac{\sigma(q^2)-1}{\sigma(q^2)+1})+2\right\}$
will be denoted as $R_5$. The numerical value of $\bar{l_5}$ was found
in Ref.[6] from the data on $\pi\rightarrow e\nu\gamma$
decay: 
\be
\bar{l_5}=13.9 \pm1.3
\ee
The absolute value of $\delta$ does not exceed 3
and it is much less than $\bar{l_5}$ but the corrections due to
 $\bar{l_5}$ and  $\bar{l_6}$ have a different sign and almost equal
values and therefore the account of $\delta$ is necessary. 

 Substituting these effective vertices into all zero order diagrams
and collecting the terms proportional to $1/F^2$ we get the final results
for ChPT corrections to structure functions $F^L_2$ and
$F^L_1$ of longitudinal photon: 

$$ F^{L(ChPT)}_2=\frac{R_6}{48\pi^2F^2}(-Q^2+p^2)F^{L(fig.1a)}_2
-\frac{R_6Q^2}{48\pi^2F^2}F^{L(fig.1b)}_2+$$

\be
+\frac{\alpha\chi}{2\pi}
\frac{R_5+\delta-R_6}{12\pi^2F^2}p^2x^2[4-4x+(1-2x)L]
\ee

and

\be
F^{L(ChPT)}_1=-\frac{\alpha\chi}{2\pi}\frac{R_6p^2}{48\pi^2F^2}
x(1-x)(1-2x)^2(L+3)
\ee
where $F^{L(fig.1a)}_2$ and $F^{L(fig.1b)}_L$
correspond to formulas ($\ref{F21a}$) and ($\ref{F21b}$). 

\section{Discussion}

$\quad$ Collecting all terms we can write final results for structure
functions $F^L_2$:

$$F^L_2=[1+\frac{R_6}{48\pi^2F^2}(-Q^2+p^2)]F^{L(fig.1a)}_2+
[1-\frac{R_6Q^2}{48\pi^2F^2}]F^{L(fig.1b)}_2+$$
\be
\label{F2f}
+\frac{\alpha\chi}{2\pi}\frac{R_5+\delta-R_6}{12\pi^2F^2}p^2x^2[4-4x+(1-2x)L]
\ee
and for $F^L_1$:
\be
\label{F1f}
F^L_1=\frac{\alpha\chi}{2\pi}\frac{p^2}{Q^2}(1-\frac{R_6Q^2}{48\pi^2F^2})
x(1-x)(1-2x)^2(L+3)
\ee

At $Q^2\approx 0.1-0.2$GeV$^2$ the large first order ChPT correction comes 
from the factors in the square brackets in front of the first terms
in the r.h.s of eq.'s ($\ref{F2f}$). These factors
have the meaning of the squares of pion formfactors in the vertices 
of the diagrams Fig. 1. Therefore, the accuracy of
eq. ($\ref{F2f}$) may be improved, if these
factors would be represented in the standard form of pionic formfactors
$(1+R_6Q^2/96\pi^2 F^2_{\pi})^{-2}$. In the numerical calculations
we use such a procedure.

 The numerical results of the calculation at few values of 
parameters are represented in Fig.~5. In the Fig. 5 one
can see the $p^2$-dependence for $F^L_2$ function.
 Since the pion mass is rather small we could expect
before calculations the following behaviour the structure functions at
low $p^2$: in the region $|p^2| \ga 0.1$Gev$^2$ the pion mass is 
inessential and structure functions have to be close to the constants.
In the region of less $p^2$ the pion mass in the Born terms
starts to work and structure functions have to go to zero sharply.

 In fact the situation is different from this expectations. After 
adding to the Born term the ChPT corrections the 
structure functions become very close to straight lines and 
one can not observe the peculiar behaviour of these function at low 
$p^2$. ChPT corrections have negative sign and result to
 decreasing of $F^L_2$ at
all $p^2$ and $x$. 

 The x-dependence of $F^L_2$ structure function is
represented in Fig. 6. The existance of $(1-2x)$ factor
 strongly supresses 
the zero order term in the $p^2/Q^2$ expansion, so the terms
 $\sim p^2/Q^2$ are important here. 
 
 The $Q^2$ dependence of $F^L_2$ is plotted at the Fig. 7. The decreasing
of $F^L_2$ with increasing of $Q^2$ is caused mainly by the factor in
the square bracket at the first and second terms in the r.h.s. 
in eq. ($\ref{F2f}$),
corresponding to the pionic formfactor at the vertices of the Fig.~1
diagrams. At Fig. 8 the $p^2$-dependece for the structure function 
$F^L_1$, determined 
by eq. ($\ref{F1f}$) is plotted.
It is seen, that $F^L_1$ is about order of magnitude less 
than $F^L_2$. In this aspect the 
photon structure function at low virtualities strongly differs from
ones at high $Q^2$. (Recall, that at high $Q^2$, $F_1=F_2/2x$.)
The small values of $F^L_1$ can be explained by absence of zero order 
term in the $p^2/Q^2$ expansion and strong numerical compensation
(accidently, the numerical value of $L\approx -3$).
For this reason the next order corrections can be important and
the results for $F^L_1$ are less reliable than for $F^L_2$.

 Let us now discuss the accuracy of the obtained results.
Since only two terms in ChPT were calculated, only the general arguments,
refering to the convergence of ChPT can be used. According to these
 arguments the expansion parameters are the ratios
of all invariants entering in the problem - $Q^2$, $|p^2|$ and $s$,
to the characteristic ChPT scale $M^2\simeq0.6$GeV$^2$. In our case
at $Q^2\simeq0.1-0.2$GeV$^2$, $|p^2|\la0.1$GeV$^2$ and this ratios
are of order $1/5-1/3$.

 The total energy s enters only in the correction $\delta$
to the $\bar{l_5}$ due to four pion interaction and at $x\ga0.15$
does not contribute to the structure functions more than $40\%$.
In the case of $p^2/Q^2$ corrections one can find that
in almost all cases the real expansion
 parameter
is $p^2/\nu$. 
So, one may expect that the accuracy of obtained results is about
$30-50\%$. The accuracy is better at intermediate $x\sim0.2-0.3$ and worse
at $x\approx 0.15$ because the ratio $s/M^2$ becames large. For this reason
there is no confidence in the results at $x<0.15$. 

 The comparision of the obtained results with experiment would be
very important. The considered above phenomena is a new field in the
application of ChPT and the comparision of the theory with experiment
in this field can shed more light on ChPT, its applicability domain,
the role of higher order ChPT corrections, etc.

 Experimentally, the photon structure functions at low virtualities
can be studied in several ways. One possibility could be the experiments
of $2\pi$ production at high flux $e^+e^-$ colliders ($\phi$-factories):
\be
\label{first}
e^+e^-\rightarrow e^+e^- + \pi^+ \pi^-.
\ee
The another way could be the 2 pion production in electron 
scattering on heavy nucleous:
\be
\label{second}
e+Z \rightarrow e+Z+ \pi^+ \pi^-.
\ee

If the momentum transfer to the nucleous is very small, less than
the inverse nucleous radius, $\sqrt{|p^2|}<1/R \simeq 50$MeV for 
intermediate
nuclei ($Z\sim 50$), than the scattering on the  nucleous is elastic and the 
suppression factor in comparision with pion electroproduction is 
$(Z\alpha)^2/Zf$, where f characterizes the suppression of two pion
electroproduction at low energies.
 In this case one of colliding gammas is Coulombic,
i.e. it is longitudinal. Since the cross section of $\gamma\gamma$
collision in this domain of $p^2$ is proportional to $p^2$ and it is 
multiplied by the square of Coulombic photon propagator 
$\propto 1/p^4$, one may expect a strong enhancement of this effect
at low $p^2$. May be there are some chances to measure the process 
($\ref{second}$) in the case, when the final nucleous is exited, or
even spallation process occures. The enhancement of the effect at
small $p^2$ will also persists here. But in this case in order to
have the proper normalization, the observation of the process
($\ref{second}$)
must be accompanied by the measurement of electroexitation of nucleous
$$e+Z \rightarrow e+Z^*$$

 The third possible way is the observation of $ \pi^+ \pi^-$ production
in $ep$ collisions:
\be
\label{third}
e+p\rightarrow e+p +\pi^+ \pi^-,
\ee
but this production of two pions in $\gamma\gamma$ collisions
is hidden in the large background, arising from usual electroproduction.
In order to get rid of this background it is necessary to separate 
the events with very small momentum transfer to the proton. 

\vspace{2cm}

\centerline{\bf Acknowledgements}

One of the authors (B.I.) is very indepted to Prof. J. Speth for his
hospitality at KFA Julich, where the part of this investigation was done
and to A. von Humboldt Foundation for the award, which gave him the
possibility to stay at KFA Julich. I.S grateful to H.Leutwyler for 
the hospitality at Bern University.  This work was supported in part
by the International Science Foundation Grant M9H300, 
by the Schweizerischer Nationalfonds
 and 
by International Association for the Promotion of Cooperation with
Scientists from the Independent States of the Former Soviet Union
 Grant INTAS-93-283.

\newpage
\centerline{\bf Figure Captions}

\bigskip

 Fig.1 - The forward $\gamma\gamma$ scattering amplitude box diagrams
 in scalar electrodynamics-the zero order of ChPT; the solid lines 
 correspond to pions, the wavy lines to photons, a) direct diagram;
b) crossing diagram.

\bigskip

Fig.2 - The diagrams of scalar electrodynamics for the  forward 
$\gamma\gamma$ scattering, arising from the term $A_{\mu}^2\phi^+\phi$
in the Lagrangian.

\bigskip

Fig.3 - The diagrams for effective $\gamma\pi\pi$ vertex, a) the diagram
corresponding to scalar electrodynamics; b) the diagram arising from
$L_6$ term in chiral Lagrangian; c) the loop diagram corresponding to
four pion interaction.

\bigskip

Fig.4 - The diagrams for effective $\gamma\gamma\pi\pi$ vertex,
a) the diagram coming from scalar electrodynamics; b), c) the diagrams
corresponding to $L_6$ and $L_5$ terms in chiral Lagrangian,
respectively; d), e), f) the loop diagrams.

\bigskip

Fig.5 - The structure function $F^L_2$ as a function of $p^2$
at fixed $Q^2=0.15$GeV$^2$, $x=0.25$ and $x=0.4$.
 The dashed line represents the 
contribution of the scalar electrodynamics(Born) 
 term at $Q^2=0.15$GeV$^2$ and $x=0.25$.

\bigskip

Fig.6 - The structure function $F^L_2$ as a function of $x$
at fixed $Q^2=0.15$GeV$^2$, $p^2=~-0.03$GeV$^2$ and $p^2=-0.07$GeV$^2$ 
 The dashed line represents 
the contribution of the Born
 term at $Q^2=0.15$GeV$^2$ and 
$p^2=-0.03$GeV$^2$.

\bigskip

Fig.7 - The structure function $F^L_2$ as a function of $Q^2$
at fixed $p^2=-0.03$GeV$^2$, $x=0.2$ and $x=0.3$.

\bigskip

Fig.8 - The structure function $F^L_1$ as a function of $p^2$
at fixed $Q^2=0.15$GeV$^2$, $x=0.15$ and $x=0.2$.

\newpage

\begin{figure}[b]
\epsfxsize=15cm
\centerline{\epsffile{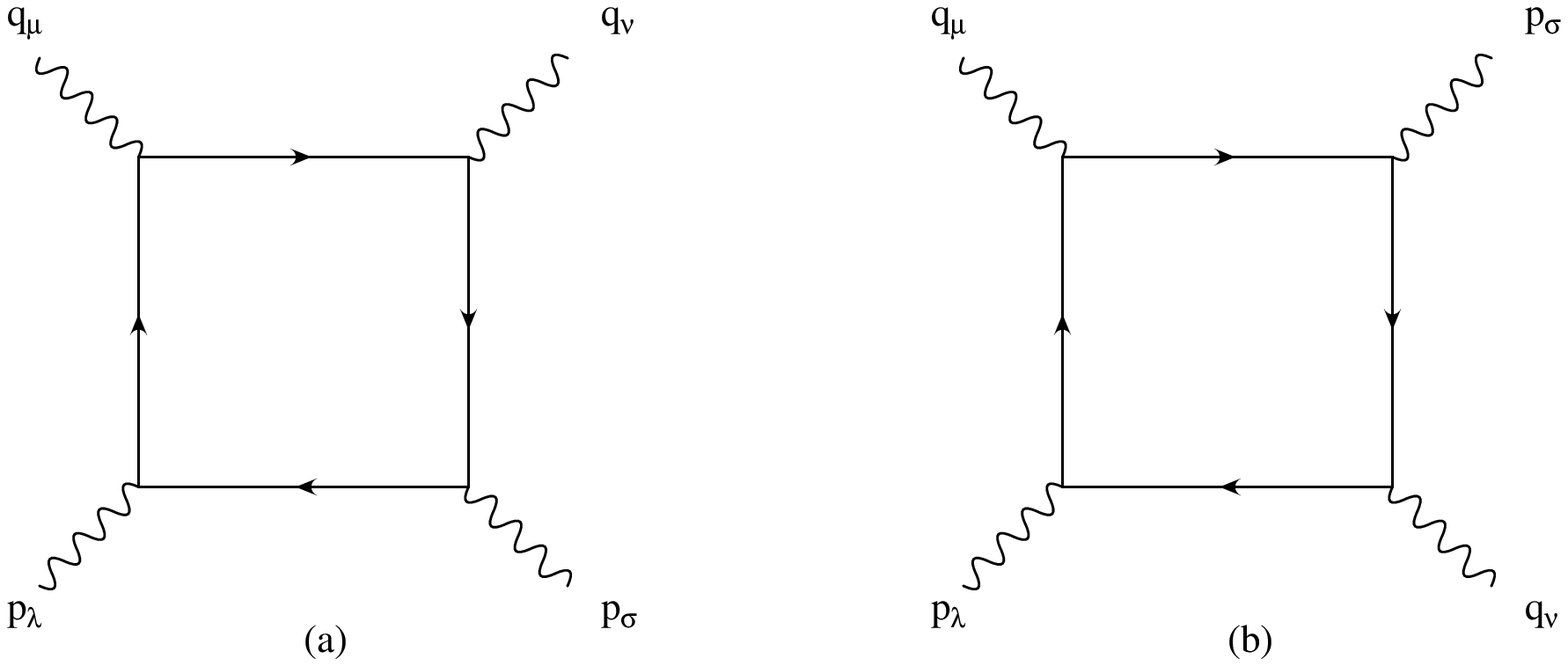}}
 \caption{}
\end{figure}

\begin{figure}[h]
\epsfxsize=15cm
\centerline{\epsffile{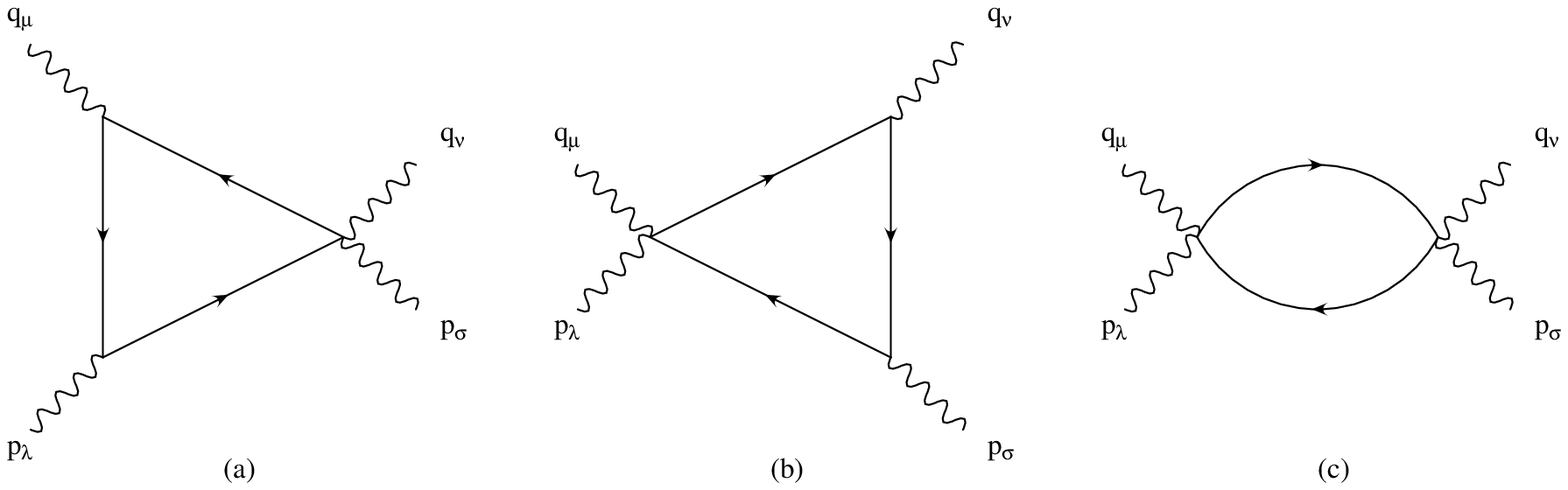}}
 \caption{}
\end{figure}

\clearpage

\begin{figure}[b]
\epsfxsize=15cm
\centerline{\epsffile{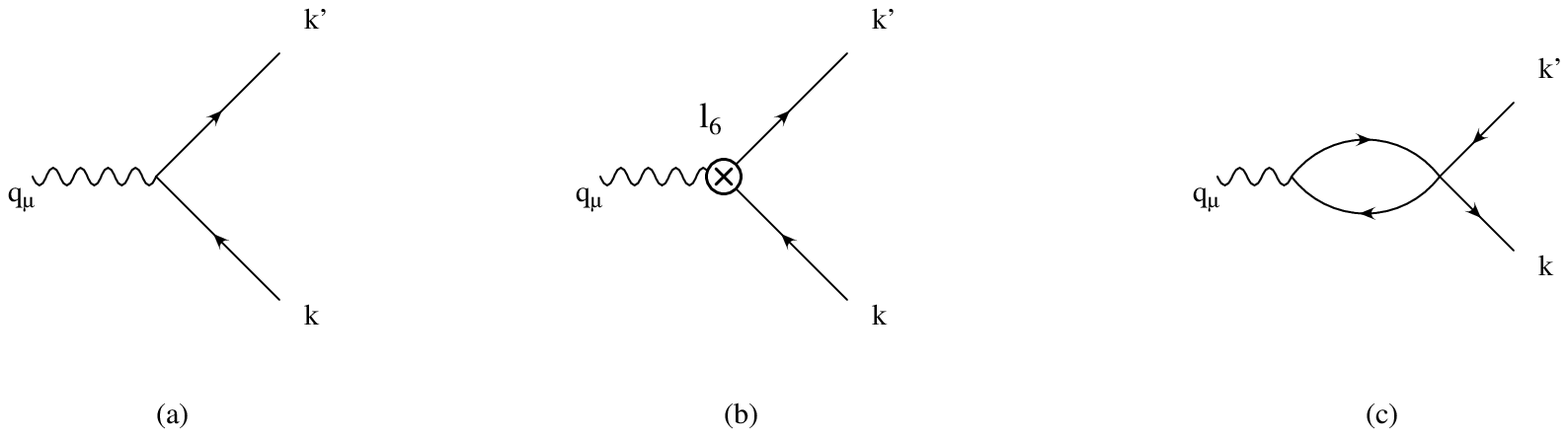}}
 \caption{}
\end{figure}
  
\begin{figure}[h]
\epsfxsize=15cm
\centerline{\epsffile{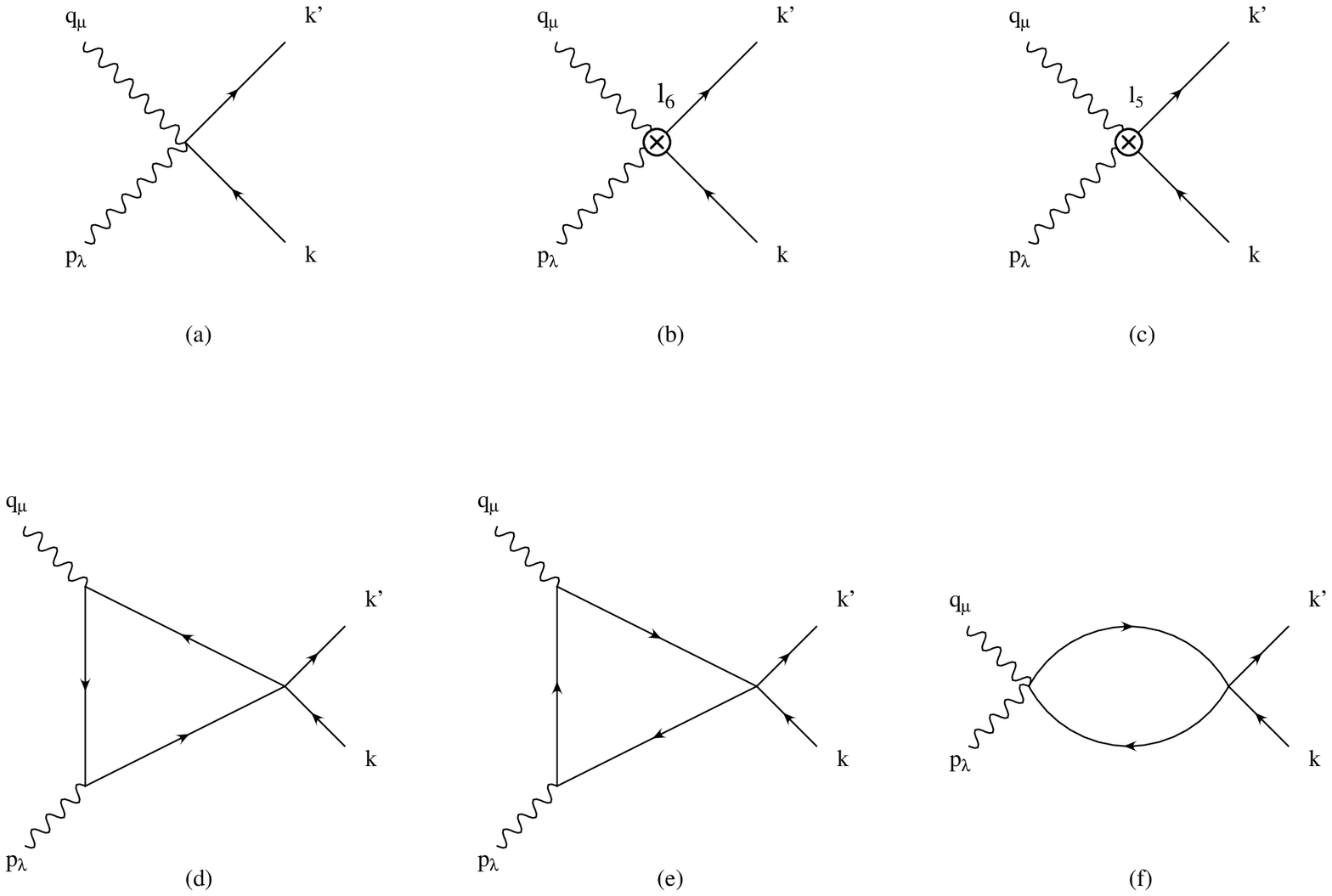}}
 \caption{}
\end{figure}

\newpage

\begin{figure}[b]
\epsfxsize=15cm
\centerline{\epsffile{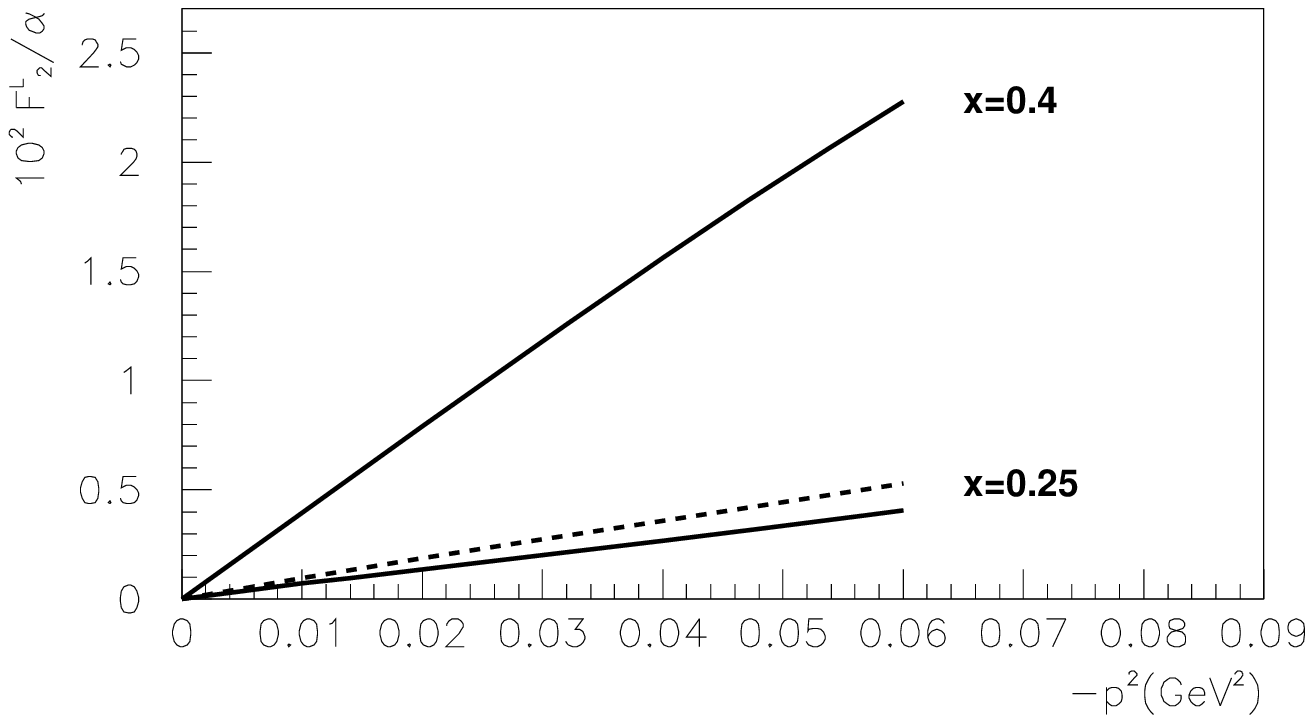}}
 \caption{}
\end{figure}

\begin{figure}[b]
\epsfxsize=15cm
\centerline{\epsffile{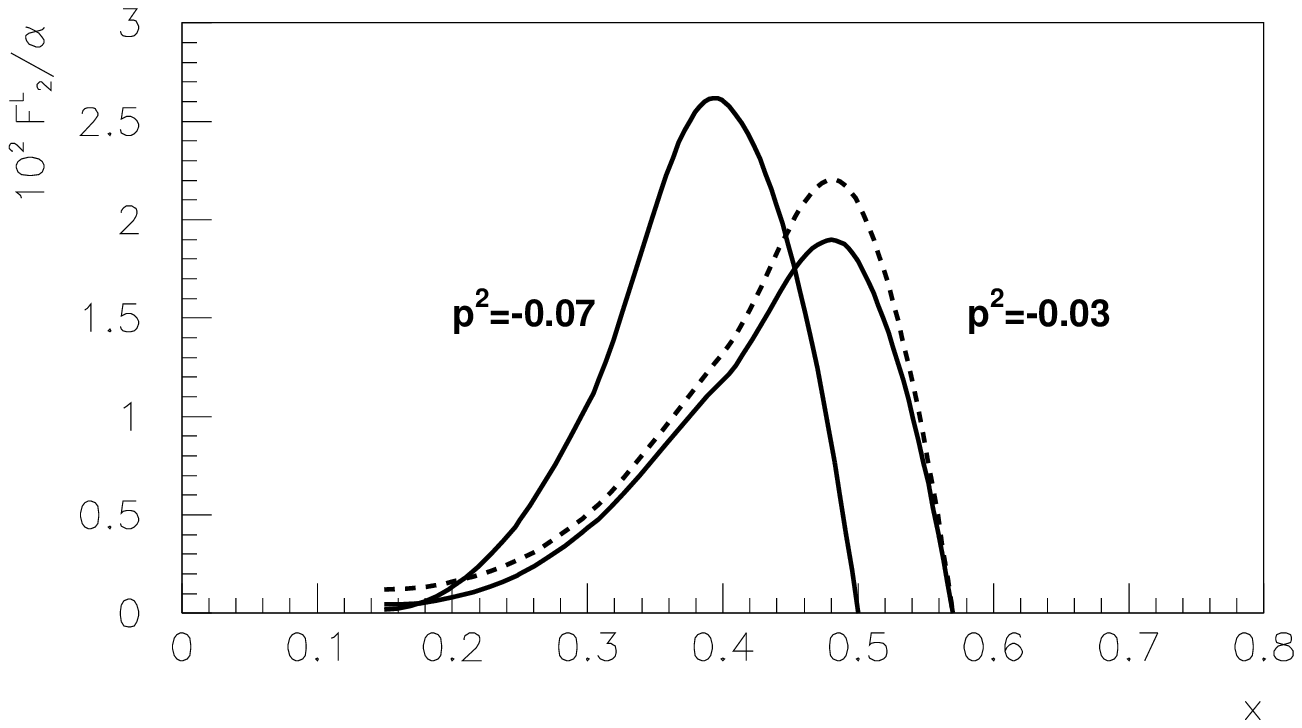}}
 \caption{}
\end{figure}

\newpage

\begin{figure}[b]
\epsfxsize=15cm
\centerline{\epsffile{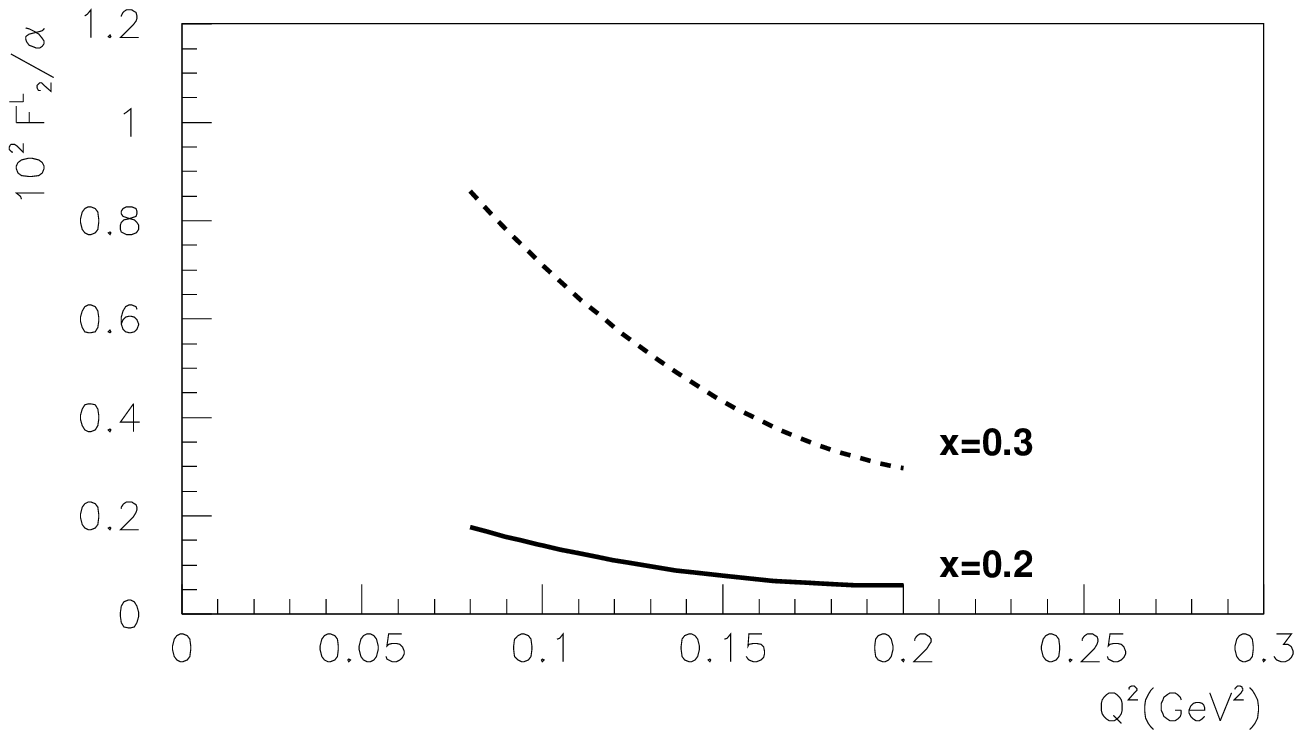}}
 \caption{}
\end{figure}

\begin{figure}[b]
\epsfxsize=15cm
\centerline{\epsffile{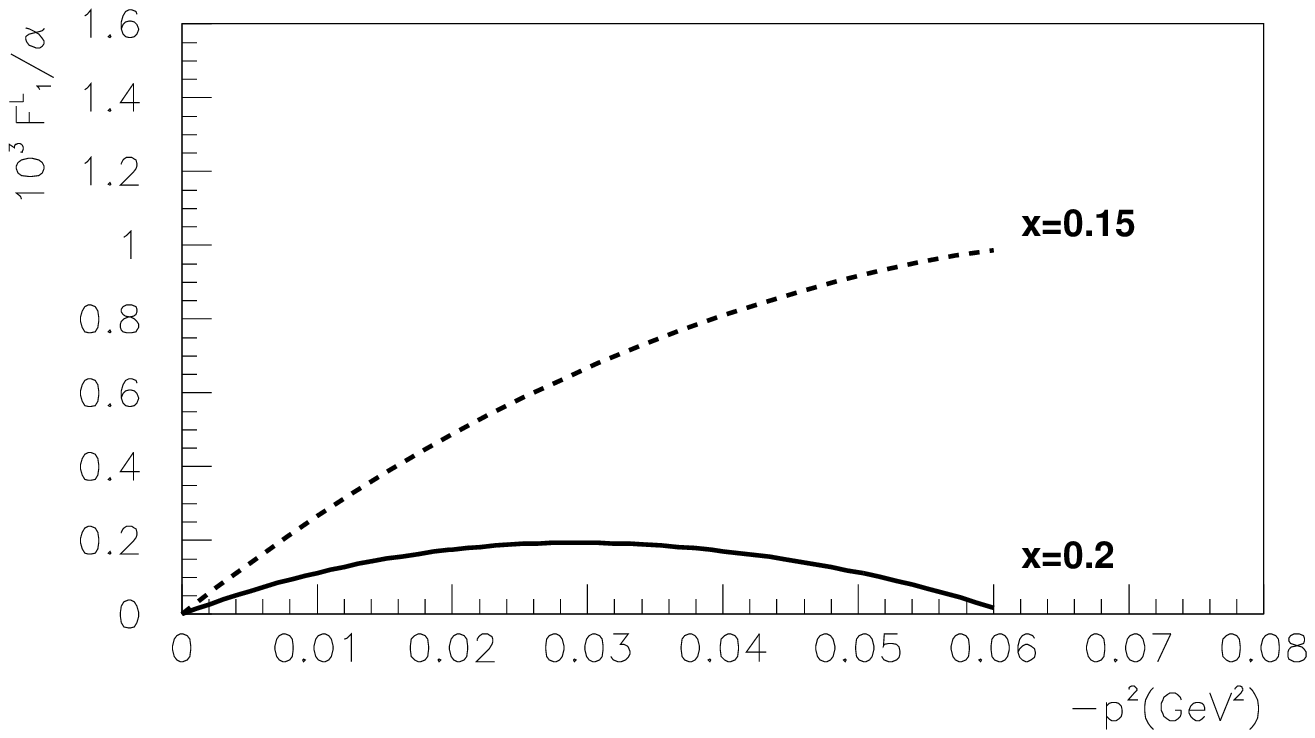}}
 \caption{}
\end{figure}

\end{document}